\documentclass[reprint,a4paper,fleqn,citeautoscript,superscriptaddress]{revtex4-1}

\usepackage{amssymb}
\usepackage{amsmath}
\usepackage{graphicx}
\usepackage[dvipsnames]{xcolor}

\usepackage[varg]{txfonts}   
\usepackage{newtxtext}
\usepackage[varg]{newtxmath}
\usepackage{textcomp}
\usepackage[version=3]{mhchem}

\usepackage{siunitx}
\usepackage[colorlinks=false,
            bookmarks=true,
            bookmarksnumbered=true]{hyperref}
\hypersetup{pdfborder = {0 0 0}}

\newcommand{\etal}{\textit{et al.}}

\csname endofdump\endcsname

\begin{document}

\preprint{\today}

\title{Critical mode and band-gap-controlled bipolar thermoelectric properties of SnSe}

\author{I. Loa}
\email[Corresponding author:~E-mail~]{I.Loa@ed.ac.uk}
\affiliation{SUPA, School of Physics and Astronomy, and Centre for Science
at Extreme Conditions, The University of Edinburgh, Edinburgh, EH9 3FD, UK}

\author{S. R. Popuri}
\affiliation{Institute of Chemical Sciences and Centre for Advanced Energy Storage and Recovery, School of Engineering and Physical
Sciences, Heriot-Watt University, Edinburgh, EH14 4AS, UK}

\author{A. D. Fortes}
\affiliation{ISIS Facility, STFC Rutherford Appleton Laboratory, Harwell Oxford, Didcot, Oxon, OX11 0QX, UK}

\author{J. W. G. Bos}
\email[E-mail~]{j.w.g.bos@hw.ac.uk}
\affiliation{Institute of Chemical Sciences and Centre for Advanced Energy Storage and Recovery, School of Engineering and Physical
Sciences, Heriot-Watt University, Edinburgh, EH14 4AS, UK}

\date{\today}

\begin{abstract}
The reliable calculation of electronic structures and understanding of
electrical properties depends on an accurate model of the crystal structure.
Here, we have reinvestigated the crystal structure of the high-$zT$
thermoelectric material tin selenide, SnSe,  between 4 and 1000~K using
high-resolution neutron powder diffraction. Symmetry analysis reveals the
presence of four active structural distortion modes, one of which is found to
be active over a relatively wide range of more than ${\pm}$200~K around the
symmetry-breaking $Pnma$--$Cmcm$ transition at 800~K. Density functional theory
calculations on the basis of the experimental structure parameters show that
the unusual, step-like temperature dependencies of the electrical transport
properties of SnSe are caused by the onset of intrinsic bipolar conductivity,
amplified and shifted to lower temperatures by a rapid reduction of the band
gap between 700 and 800 K. The calculated band gap is highly sensitive to small
out-of-plane Sn displacements observed in the diffraction experiments. SnSe
with a sufficiently controlled acceptor concentration is predicted to produce
simultaneously a large positive and a large negative Seebeck effect along
different crystal directions.
\end{abstract}

\maketitle


\section{Introduction}

Recently, single crystals of the binary compound tin selenide, SnSe, were
reported to exhibit exceptionally good thermoelectric properties
\cite{ZLZS14,CWHP18,CSZZ18}. The discovery of high thermoelectric
efficiency in a {``}simple{''} bulk material has added to the renewed interest
in thermoelectric materials for energy harvesting applications, where
waste heat is converted directly to electricity in reliable,
low-maintenance semiconductor devices. The high performance in SnSe arises
from the combination of an ultralow lattice thermal conductivity and a
rapid increase in the thermoelectric power factor with temperature above
600~K. The low thermal conductivity has been attributed to highly
anharmonic lattice dynamics \cite{LHMB15,BHLM16,SBPW16,HD16pv2,PPDV17},
whereas the unusual, almost step-like changes of the electrical
conductivity and Seebeck coefficient around 700~K have remained largely
unexplained
\cite{KWKT15,SK15,GWKH15,DGY15a,DHBR16,HLZY15,XZYM16,YZYW15,MUOK17}.

\begin{figure}[bt]
   \centering
   \includegraphics[width=86mm]{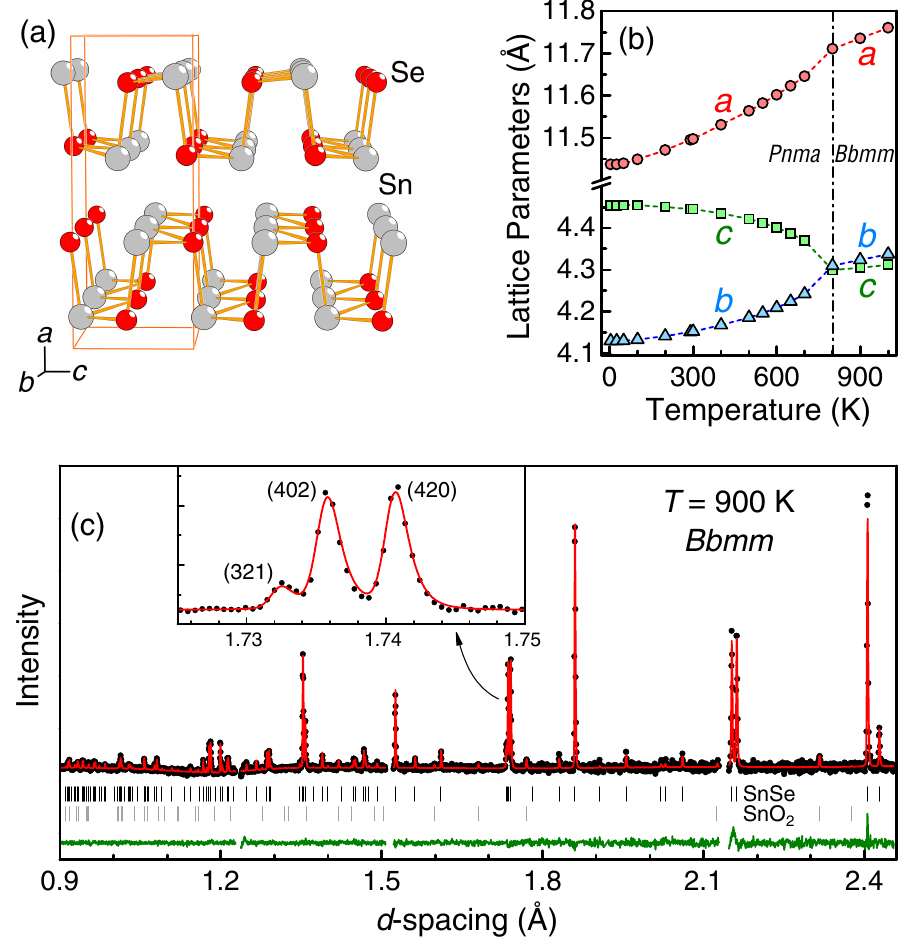}
   \caption{(a) Crystal structure of SnSe below 800~K in the $Pnma$
            setting. %
            (b)~Temperature dependence of the lattice parameters of SnSe. %
            (c)~Neutron powder diffraction pattern of SnSe at 900~K ($Bbmm$ phase)
            and Rietveld fit. Measured data are shown by black circles, the Rietveld
            fit by red lines, and the difference curve by the green line at the
            bottom. Tick marks show the fitted peak positions for SnSe (black, upper
            row) and a $\sim$1wt\% SnO$_{2}$ impurity that was also observed (gray,
            lower row). Three small regions in $d$-spacing with artifacts from the
            TOF detection were excluded from the fits.}
     \label{fig:SnSe_structure_lattparam_diffpatt}
\end{figure}

The crystal structure of SnSe can be described as a highly distorted rock salt derivate
\cite{WC79,SW81,CPS86}. Above 800~K, the structure has space group symmetry $Cmcm$ and
features rock-salt-type double layers, see
Figs.~\ref{fig:SnSe_structure_lattparam_diffpatt}(a) and \ref{fig:distortion_modes}(a).
Upon cooling below 800~K, the structure undergoes a displacive phase transition to lower
symmetry, $Pnma$. Instead of space group $Cmcm$, we will use henceforth the non-standard
setting $Bbmm$, so as to keep the crystal axes in the same orientation as for $Pnma$
\cite{LHDP15}. The electronic structure and transport properties of SnSe were
investigated in several computational studies in the framework of density functional
theory (DFT) \cite{KWKT15,SK15,GWKH15,DGY15a,DHBR16,HLZY15,XZYM16,YZYW15,MUOK17},
revealing a complex arrangement of anisotropic bands that contribute to the electrical
transport. Surprisingly, none of these studies considered the gradual variation of the
crystal structure with temperature. Most commonly, only the crystal structures in the
low- and high-temperature limit (typically $\sim$300~K and $\sim$800~K) were used.
Although much insight has been gained from these investigations, a calculation of the
electronic structure and transport properties directly from the temperature-dependent
crystal structure data is missing.

Here, we present a reinvestigation of the crystal structure of SnSe, extending
the temperature range to \mbox{4--1000~K}, using high-resolution neutron powder
diffraction and symmetry mode analysis to determine the active structural order
parameters. This structural information is then used to follow the evolution of
the electronic structure of SnSe as a function of temperature using DFT
calculations. These reveal a direct correlation between the primary order
parameter and the electronic band gap; they show that small atomic
displacements have a profound effect on the electronic structure, and they
provide a coherent explanation for the unusual temperature dependence of the
experimentally observed transport properties. We will also discuss the
possibility of realizing simultaneously a large positive and negative Seebeck
effect along different directions of a single crystal.


\section{Techniques}

SnSe was synthesized as described previously \cite{PPDM16}.  Neutron powder diffraction
data were collected on a \mbox{10-gram} sample on the time-of-flight (TOF) High
Resolution Powder Diffractometer (HRPD), equipped with a He-flow cryostat (4--290~K) and
furnace (300--1000~K), at the ISIS spallation neutron source, Rutherford Appleton
Laboratory, UK. High-temperature patterns were collected in a vacuum-sealed
(\SI{E-5}{bar}) quartz tube to avoid sublimation of SnSe at high temperatures. The
diffraction patterns are not affected by texturing and enable the determination of
accurate atomic coordinates and thermal displacement parameters. Structural refinements
using the Rietveld method were carried out with the GSAS program
\cite{soft:GSAS,Tob01,SupplMat}.

Electronic structure calculations were performed in the framework of density functional
theory (DFT) and the full-potential augmented-plane-wave~+~local orbital (APW+lo)
approach as implemented in the \textsc{\small Wien2k} code \cite{soft:Wien2k,SupplMat}.
Exchange and correlation effects were treated with the revised generalized gradient
approximation for solids (PBEsol) \cite{PRCV08} in the total-energy calculations, and the
modified Becke-Johnson potential by Tran and Blaha \cite{TB09} was used in the band
structure and band gap calculations. Spin-orbit coupling was found to have a negligible
effect on the band structure and was therefore not included. From the electronic band
structures, electrical transport properties were calculated in the framework of Boltzmann
transport theory as implemented in the \textsc{\small BoltzTraP} code
\cite{MS06,SupplMat}.

\section{Results and Discussion}
\subsection{Crystal structure}

The structural parameters of SnSe in the temperature range of 4--1000~K were
obtained from Rietveld refinements.
Figure~\ref{fig:SnSe_structure_lattparam_diffpatt}(c) shows an example of a
diffraction pattern and Rietveld fit, and more extensive data are presented in
the Supplementary Material \cite{SupplMat}, including all structural parameters
in tabulated form. The temperature dependence of the lattice parameters
[Fig.~\ref{fig:SnSe_structure_lattparam_diffpatt}(b)] is in line with previous
results \cite{WC79,SW81,CPS86,APJF98,SZI16,SNDF16}. All axes show gradual
changes up to $\sim$600~K, followed by more rapid changes up to the phase
transition at $\sim$800~K. The $a$ and $c$ parameter approach each other with
increasing temperature, and they are reversed in length from 800~K onwards. A
noteworthy detail is illustrated in the inset of
Fig.~\ref{fig:SnSe_structure_lattparam_diffpatt}(c): The doublet of the (402)
and (420) reflections of the high-temperature phase is clearly split, thereby
demonstrating unambiguously that the metric remains orthorhombic, contrary to a
previous report \cite{APJF98}. The evolution of the positional parameters with
temperature (see Supplementary Material \cite{SupplMat}) confirms the
symmetry-raising transition from the $Pnma$ to the $Bbmm$ ($Cmcm$) space group
symmetry at $\sim$800~K \cite{WC79,SW81,CPS86}. Our refined structural
parameters are consistent with those reported recently by Sist \etal\
\cite{SZI16} (x-ray powder diffraction) and Serrano-S\'{a}nchez \etal\
\cite{SNDF16} (neutron powder diffraction) but covering a wider temperature
range (Fig.~S4 in the Supplementary Material \cite{SupplMat}).

\begin{figure}[t]
     \centering
     \includegraphics[width=86mm]{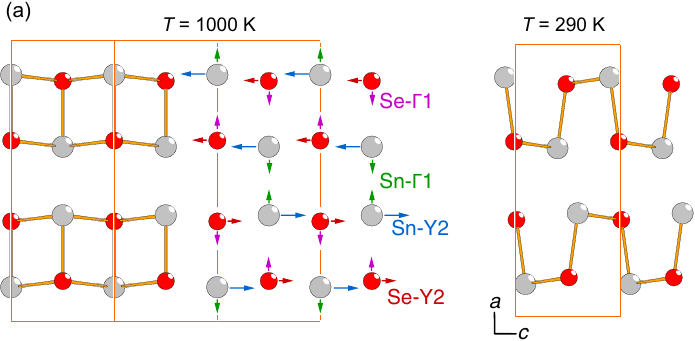}\\[3ex]
     \includegraphics[width=86mm]{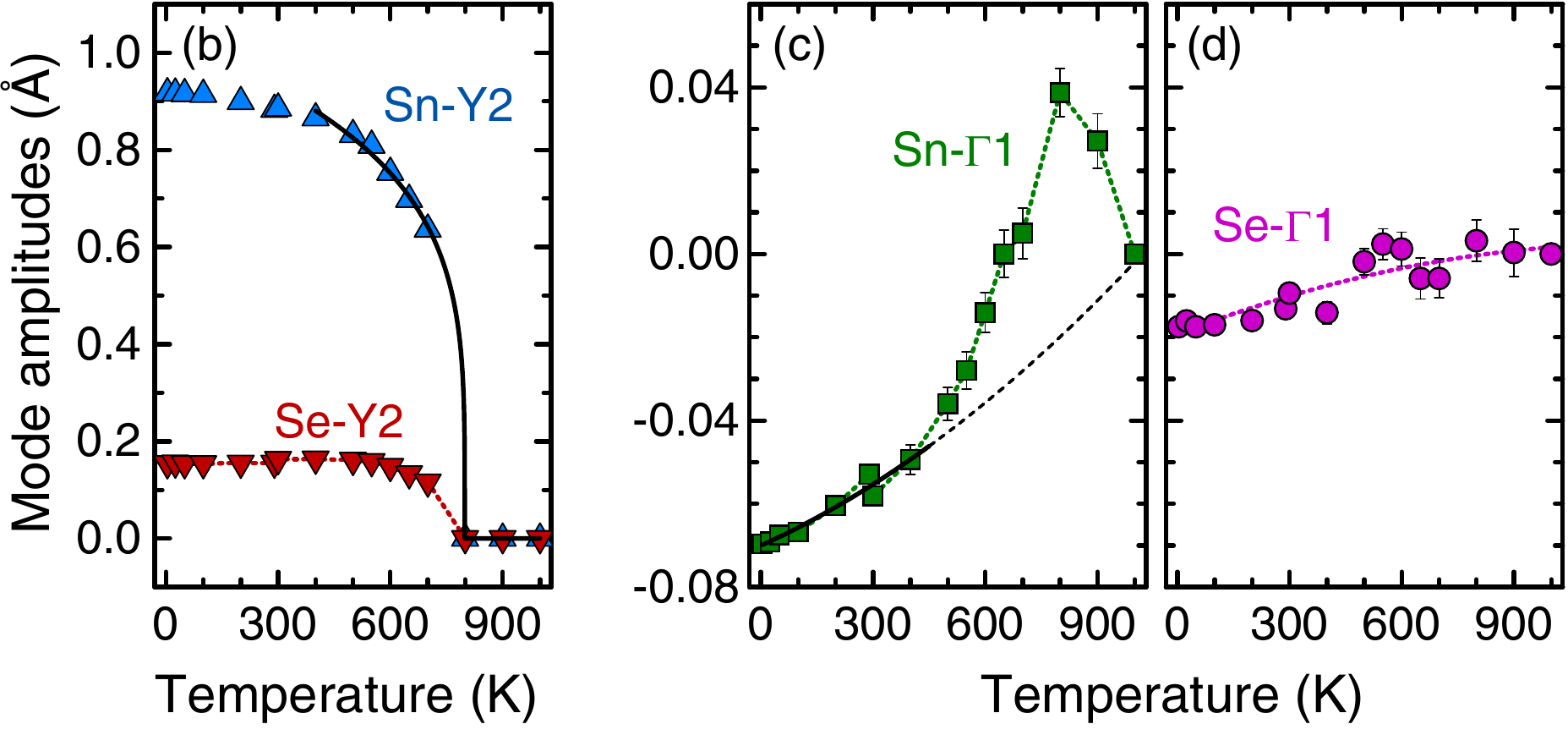}
     \caption{(a) Atomic displacement patterns of the active
              symmetry-adapted distortion modes of SnSe. The arrows are not
              to scale. (b--d) Amplitudes of the distortion modes as a
              function of temperature, see text for details. Solid black
              lines in (b,c) are fits to the data as explained in the text,
              the dashed line in (c) shows a quadratic extrapolation of the
              4--400-K data, and the dotted lines in (b--d) are guides to
              the eye.}
     \label{fig:distortion_modes}
\end{figure}

In order to characterize the phase transition further, we found it helpful to
consider not the bare fractional atomic coordinates, but instead the
\emph{symmetry-allowed distortion modes} \cite{CSTH06}. These describe the
shifts of the equilibrium atomic positions relative to the high-temperature,
high-symmetry structure. The amplitudes of these distortions were determined as
a function of temperature from the refined structure parameters using the
\textsc{Isodistort} software \cite{CSTH06}, and the high-symmetry,
\mbox{1000-K} structure was used as the reference. By convention, each mode
amplitude is defined as the sum of the displacements of all displaced atoms in
the unit cell, added in quadrature. For all distortion modes considered here,
the mode amplitude is twice the displacement of an individual atom.

Four modes were found to be active during the $Bbmm$--$Pnma$ transition: the Y2
and $\Gamma$1 modes of both Sn and Se. The atomic displacement patterns of
these distortion modes and the variation of the fitted mode amplitudes with
temperature are shown in Fig.~\ref{fig:distortion_modes}. The Y2 modes
correspond to shifts parallel to SnSe layers (and in-phase within each double
layer), whereas the $\Gamma$1 modes describe motions perpendicular to the
layers which are anti-phase within each double layer.

By far the largest change in amplitude was observed for the Sn-Y2 mode, see
Fig.~\ref{fig:distortion_modes}(b). In the 400--1000~K range, the Sn-Y2 amplitude shows
critical behavior with a transition temperature $T_s$ of 800~K and an exponent $\beta =
0.23(1)$ (solid black line). The amplitude of the Se-Y2 distortion mode is $\sim$5 times
smaller than that of Sn-Y2. Interestingly, the Sn-$\Gamma$1 mode with atomic
displacements perpendicular to the layers exhibits a pronounced anomaly around the
transition temperature [Fig.~\ref{fig:distortion_modes}(c)]: A slightly superlinear
increase in mode amplitude from 4--400 K is followed by a rapid increase up to $T_s =
800~K$, and an equally rapid decrease at higher temperature. Quadratic extrapolation of
the low-temperature behavior reproduces the experimental \mbox{1000-K} data point,
supporting the notion of anomalous behavior from $\sim$400 up to 1000~K. For
completeness, the Se-$\Gamma$1 mode amplitude exhibits a small steady variation over the
whole 4--1000~K interval [Fig.~\ref{fig:distortion_modes}(d)].

In summary, the symmetry analysis shows that the motion of the Sn and Se atoms during the
$Pnma$--$Bbmm$ transition can be decoupled into independent, symmetry-adapted
displacements parallel and perpendicular to the SnSe double layers. It identifies the
Sn-Y2 mode as a primary order parameter for the phase transition, corresponding to a
large Sn displacement within the double layers [Fig.\ref{fig:distortion_modes}(b)]. Upon
heating, this displacement starts to become significant above 600~K and saturates at
800~K, consistent with the early work by von~Schnering and Wiedemeier \cite{SW81}.

The symmetry analysis reveals furthermore anomalous Sn displacements
perpendicular to the rocksalt layers (Sn-$\Gamma$1 mode), which have not been
recognized in previous studies \cite{SW81,CPS86,APJF98,SZI16,SNDF16}. They
occur over a wide temperature interval around the phase transition at 800~K.
This mode anomaly affects mainly the shortest Sn-Se bond distance (see the
Supplementary Material \cite{SupplMat}). We will demonstrate below that this
structural detail has a signifcant effect on the electronic properties.

\subsection{Electronic Structure}

We now turn to the DFT calculations, which were performed in order to
understand how the structural variations of SnSe with temperature affect the
electronic structure and transport properties. The modified Becke-Johnson (mBJ)
exchange and correlation potential \cite{TB09} was employed here in order to
overcome the well-known problem that DFT calculations with standard LDA and GGA
potentials significantly underestimate semiconductor band gaps. For the
ambient-temperature crystal structure, the band structure calculated here with
the mBJ potential is in close agreement with results obtained previously in GW
calculations \cite{SK15,GWKH15}, which aim specifically at describing the
electronic excitation spectrum correctly. The calculated band structure is also
consistent with photo emission data \cite{WXTJ17}. The band structure
calculations were then performed using the structural parameters measured
between 4 and 1000~K.


\begin{figure}
     \centering
     \includegraphics[width=86mm]{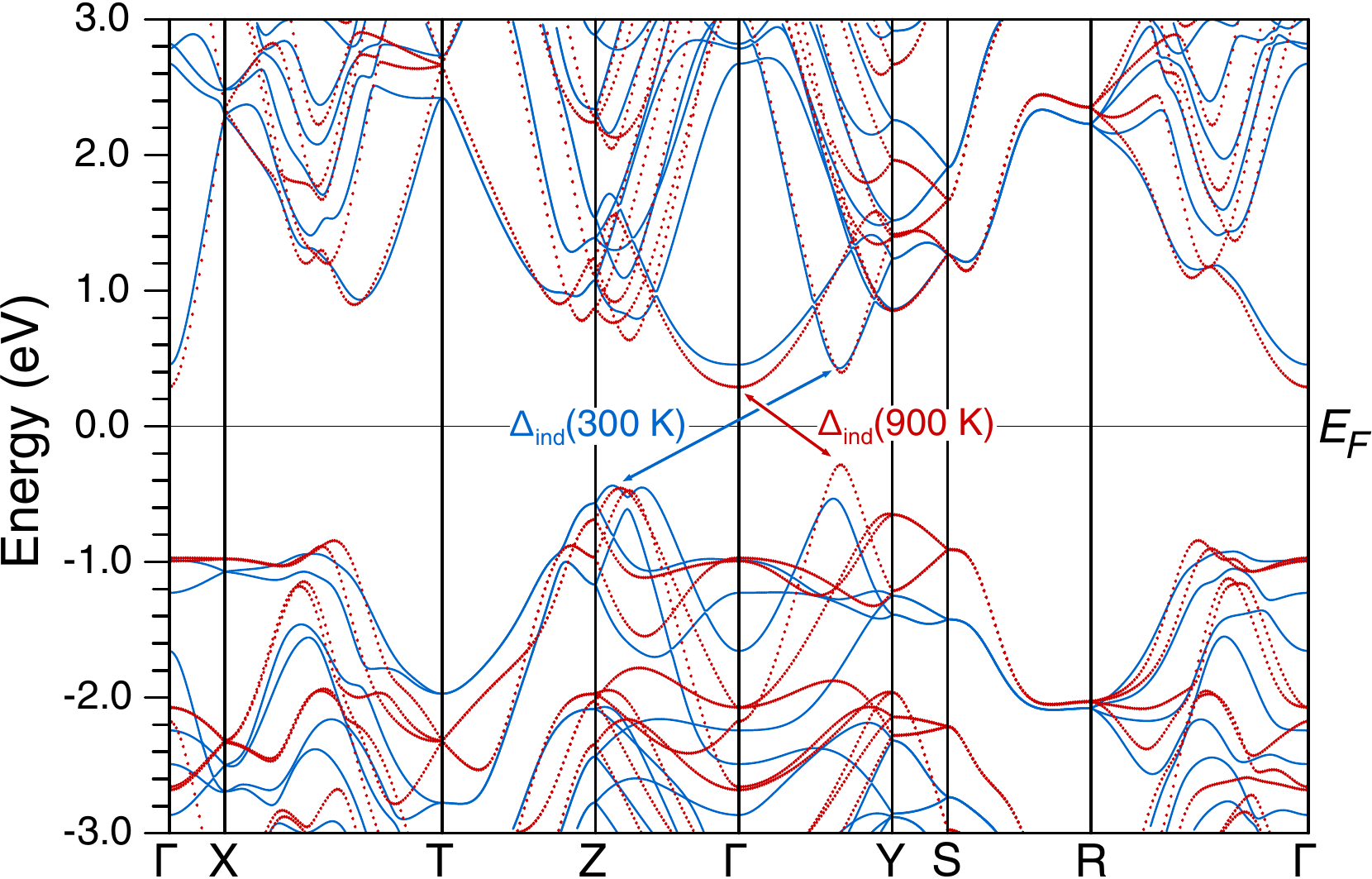}
     \caption{Calculated band structure of SnSe at 300~K (solid lines, blue) and
              900~K (dotted lines, red). The special points refer to the $Pnma$
              Brillouin zone, which was used for both calculations to allow for a
              direct comparison.}
     \label{fig:bands}
\end{figure}

The calculated band structure shows significant qualitative changes across the
$Pnma$--$Bbmm$ transition, with most of the changes occurring between 700 and
800~K. In these calculations, SnSe remains an indirect semiconductor at all
temperatures, but the locations of the valence band maximum and the conduction
band minimum change during the transition, as illustrated in
Fig.~\ref{fig:bands}. It should be noted, however, that details such as the
exact relative energies of band extrema become less relevant with increasing
temperature because the Fermi-Dirac distribution broadens, so that the
electrical transport properties result increasingly from an average over many
bands.


\begin{figure}
     \centering
     \includegraphics[width=70mm]{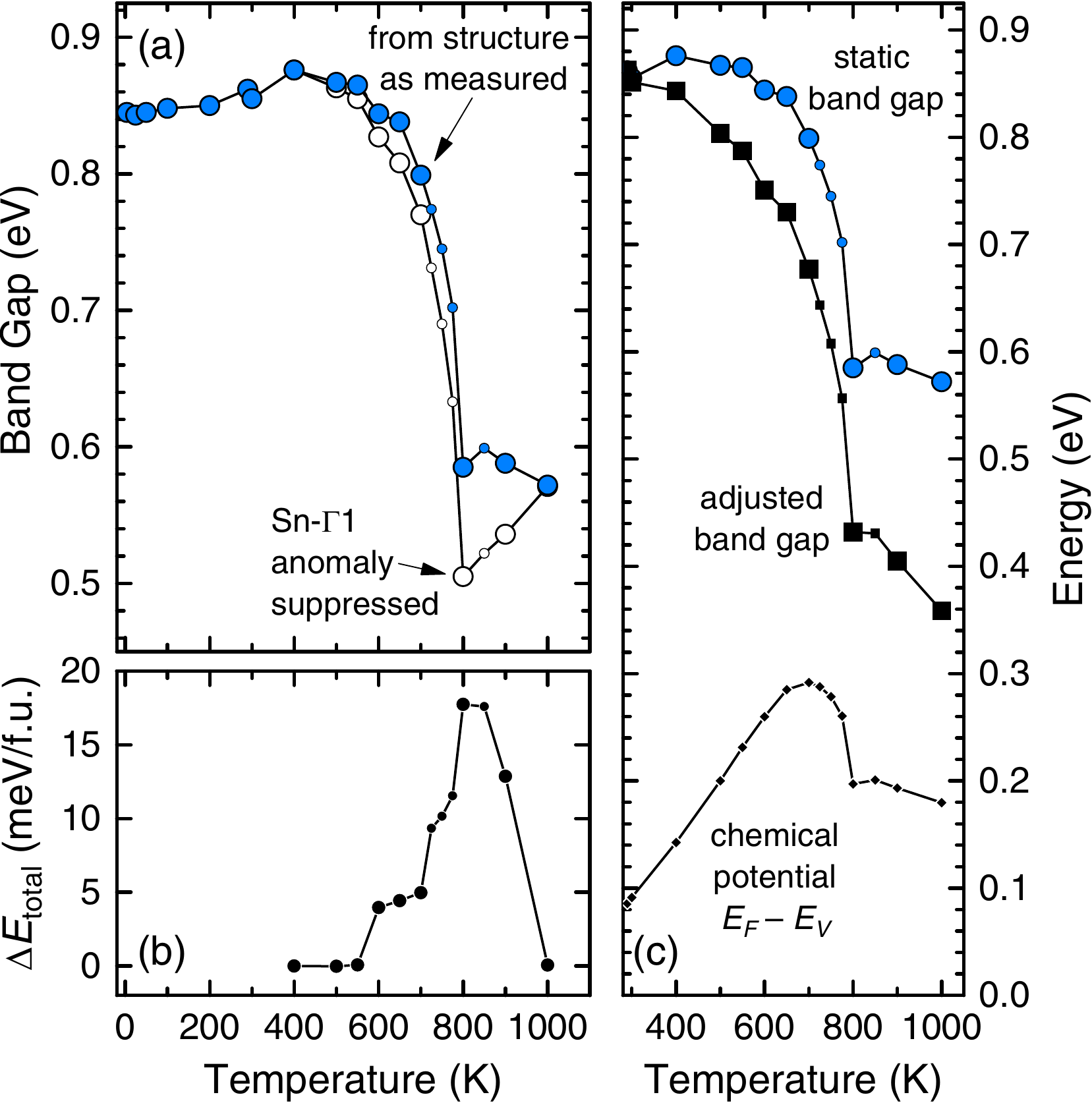}
     \caption{(a) Evolution of the calculated indirect band gap of SnSe with
              temperature. Solid symbols show results based on the crystal structure
              as measured, and open symbols where the peak of the \mbox{Sn-$\Gamma$1}
              distortion mode around 800~K was suppressed [dashed line in
              Fig.~\ref{fig:distortion_modes}(c)]. The band gap is that of the static
              lattice. Small symbols represent additional computational results where
              interpolated structural parameters were used. (b)~Difference in total
              energy between that of the hypothetical structure with the Sn-$\Gamma$1
              anomaly suppressed and that of the crystal structure as measured. (c)
              Static band gap and gap adjusted for thermal motion after \cite{PC90},
              and chemical potential $E_F$ relative to the valence band maximum,
              $E_V$, for an extrinsic hole concentration of $n_p =
              \SI{5E17}{cm^{-3}}$.}
     \label{fig:DFT_gap_dEtot}
\end{figure}

\begin{figure}
     \centering
     \includegraphics[scale=0.45]{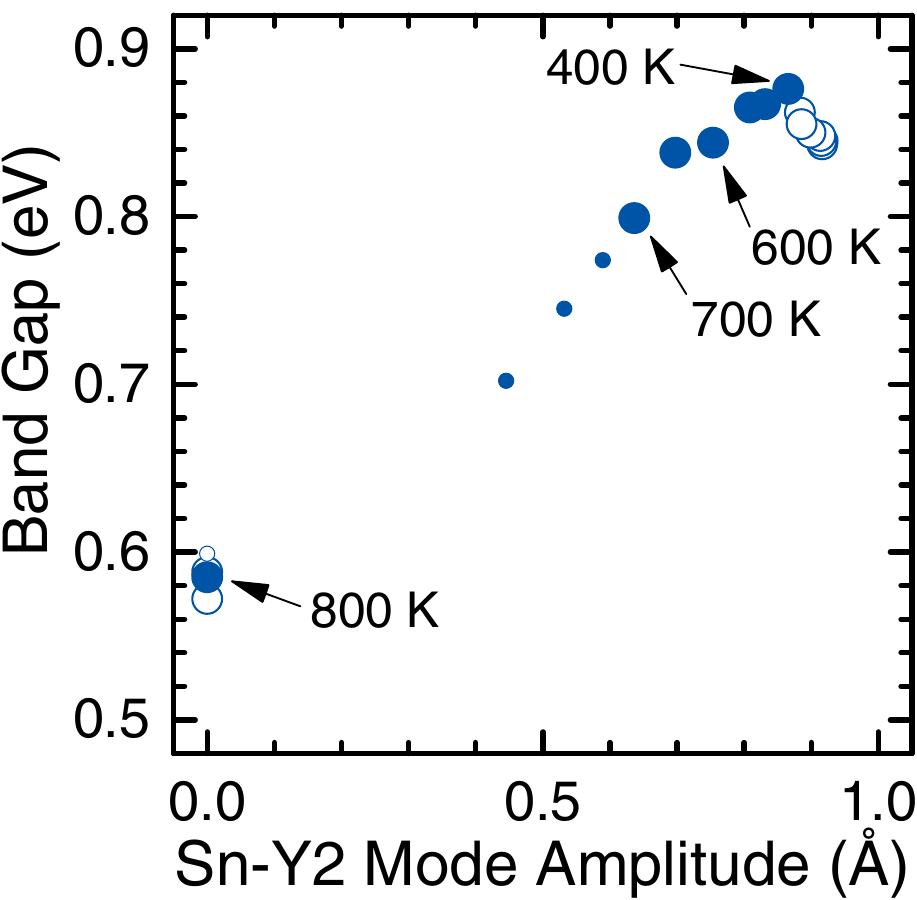}
     \caption{Correlation of the calculated band gap of SnSe with the amplitude of
              the Sn-Y2 critical mode. Open symbols mark data points below 400~K and
              above 800~K; small symbols indicate results where interpolated
              structural parameters were used in the calculations.}
     \label{fig:gap_Sn-Y2}
\end{figure}
Figure~\ref{fig:DFT_gap_dEtot}(a) shows that the calculated band gap decreases with
increasing temperature, following the primary order parameter (Fig.~\ref{fig:gap_Sn-Y2}).
In addition, the effect of the seemingly small Sn-$\Gamma$1 distortion mode on the band
gap is profound: When the anomaly of this mode around 800~K is suppressed [dashed line in
Fig.~\ref{fig:distortion_modes}(c)], a further reduction of the band gap by 80~meV at
800~K is obtained. Figure \ref{fig:DFT_gap_dEtot}(b) illustrates that suppressing the
peak in the Sn-$\Gamma$1 amplitude around 800~K also has significant effect on the total
energy: it increases by up to $\sim$18~meV per formula unit. Both effects demonstrate how
very sensitively the electronic structure responds to details of the crystal structure.

It is therefore important to bear in mind that the band structure calculations
were performed assuming a static crystal lattice, as is common practice
\cite{KWKT15,SK15,GWKH15,DGY15a,DHBR16,HLZY15,XZYM16,YZYW15,MUOK17}. However,
the thermal motion of the atoms at the high temperatures of interest for
thermoelectrics will affect the band gap and thus the transport properties. For
the following calculations of transport properties, we adjusted the calculated
band gap above room temperature by \SI{-0.30}{meV/K} based on the experimental
results by Parenteau and Carlone \cite{PC90} so as to obtain a realistic
estimate of the high-temperature band gap [Fig.~\ref{fig:DFT_gap_dEtot}(c)];
see the Supplementary Material \cite{SupplMat} for full details. A full
\textit{ab initio} calculation of the temperature dependence of the band gap is
desirable, but it would require an accurate modeling of the anharmonic lattice
dynamics \cite{LHMB15,BHLM16,SBPW16,HD16pv2} and its effect on the electronic
structure
---  a formidable task beyond the scope of this work.

\subsection{Transport Properties}

\begin{figure}[t]
     \centering
     \includegraphics[width=86mm]{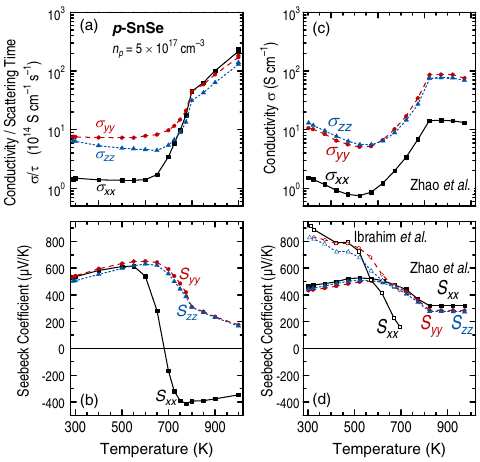}
     \caption{(a) Calculated electrical conductivity $\sigma$ divided by
              the scattering time $\tau$ of $p$-SnSe and (b) calculated
              Seebeck coefficient as a function of temperature. A
              hole-doping concentration of $n_p = \SI{5E17}{cm^{-3}}$ at
              ambient temperature and the adjusted band gaps from
              Fig.~\ref{fig:DFT_gap_dEtot}(c) were used in both cases. (c,
              d) Corresponding experimental results by Zhao \etal\
              \cite{ZLZS14} (solid symbols) and Ibrahim \etal\
              \cite{IVSC17} (open symbols).}
     \label{fig:sigma-Seebeck}
\end{figure}

Figure~\ref{fig:sigma-Seebeck}(a) shows the calculated temperature dependence
of the conductivity along the three orthorhombic crystal axes ($a,b,c \to
xx,y\!y,zz$), normalized by the electron scattering time $\tau$. The extrinsic
hole concentration was chosen as $n_p = \SI{5E17}{cm^{-3}}$ at 300~K as
suggested by Hall effect data by Zhao \etal\ \cite{ZLZS14}, and the number of
extrinsic holes per unit cell was assumed to remain constant at higher
temperatures. The calculated conductivity shows a very pronounced anisotropy in
the 300--600-K range, a steep, step-like increase in a narrow range from 600 to
800~K and a more moderate increase with temperature above 800~K. Overall, these
results compare well with the experimental results for the single-crystal
conductivity by by Zhao \etal\ \cite{ZLZS14}, see
Fig.~\ref{fig:sigma-Seebeck}(c).

These computational results  explain the unusual step-like increase in
conductivity in SnSe between 600 and 800~K as well as the marked kink at 800~K
\cite{ZLZS14}. The conductivity starts to increase along with the increased
carrier concentration at the onset of intrinsic bipolar conductivity, as
expected, but this gradual process is \emph{amplified and shifted to lower
temperatures} by the rapid reduction of the band gap between 700 and 800~K
(Fig.~\ref{fig:DFT_gap_dEtot}). The structural transition and the associated
rapid reduction of the band gap are completed at 800~K, giving rise to the
marked change in slope of the conductivity. As noted above, the calculated
conductivity is with respect to the scattering time $\tau$, which generally
decreases with increasing temperature and will hence reduce somewhat the rise
in conductivity above 800~K. A notable difference between the computational and
experimental results is the apparent disappearance of the anisotropy at high
temperature in the calculations. The calculated conductivity is normalized with
respect to the electron scattering $\tau$, and a significant difference between
the in-plane and out-of-plane electron scattering times would recover the
experimentally observed anisotropy in the conductivity.

We performed additional calculations that show that the results for the
temperature dependence of the electrical conductivity do not depend
qualitatively on details of the underlying band structures (Figs. 5(a) and 6(a)
in the Suppl.\ Mat.\ \cite{SupplMat}): (i) If the as-calculated band structures
with without band-gap adjustment on account of thermal motion are used, the
results remain qualitatively the same; the reduced change in band gap leads
only to a smaller increase in conductivity. (ii) The results also remained
largely unchanged in calculations in which the same, fixed 300-K band structure
together with the adjusted band gaps from Fig.~\ref{fig:DFT_gap_dEtot}(c) was
used. In this scenario, the effective masses and relative order of band extrema
are fixed. The key conclusion is therefore that the unusual temperature
dependence of the conductivity is controlled by the variation of the band gap
(via the associated variation of the charge carrier concentration), which in
turn correlates with the Sn-Y2 critical mode (Fig.~\ref{fig:gap_Sn-Y2}), but
not by the details of the band shapes, i.e.\ the effective masses and relative
order of band extrema.

\begin{figure}[b]
     \centering
     \includegraphics[width=70mm]{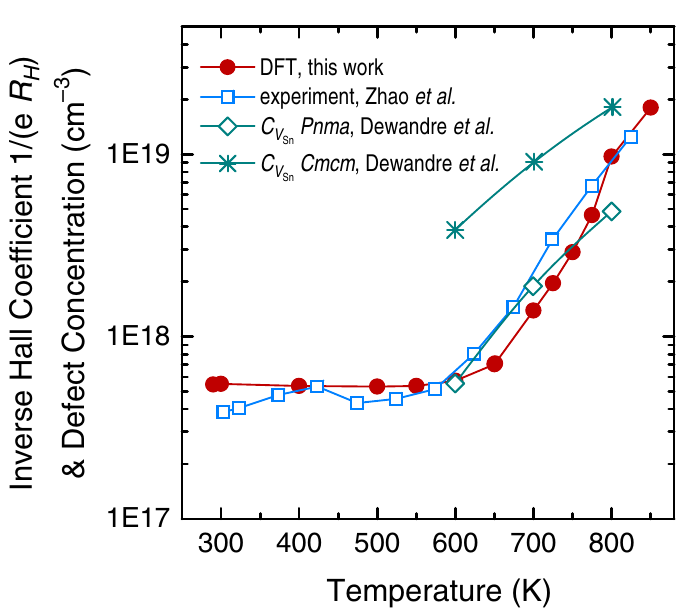}
     \caption{Inverse Hall coefficients, $1/(e R_H)$, and Sn vacancy
              concentrations, $C_{V_\text{Sn}}$, in SnSe as a function of
              temperature. Red solid circles show the average inverse Hall
              coefficient obtained here from Boltzmann transport
              calculations assuming hole-doping concentration of $n_p =
              \SI{5E17}{cm^{-3}}$ at ambient temperature and the adjusted
              band gaps from Fig.~\ref{fig:DFT_gap_dEtot}(c). Blue open
              squares show the corresponding experimental data by Zhao
              \etal\ \cite{ZLZS14}. Cyan diamonds and stars show the
              carrier concentration obtained from the Sn$^{2-}$ vacancy
              model by Dewandre \etal\ \cite{DHBR16} for the $Pnma$ and
              $Cmcm/Bbmm$ phases, respectively.}
     \label{fig:invHall}
\end{figure}

As a further test of the calculations and the above conclusions, we compare in
Fig.~\ref{fig:invHall} the average inverse Hall coefficient, $1/(e R_H)$, from
the measurements by Zhao \etal\ \cite{ZLZS14} and from our calculations. These
measures of the effective charge carrier concentration are in very good
agreement over the whole temperature range, which confirms the view that the
steep increase in conductivity follows the increase in charge carrier
concentration at the onset of intrinsic conductivity at $\sim$600~K, where the
band gap also starts to decrease.

The above conclusions are in contrast to the recently advanced notion that
the carrier concentration in SnSe is dominated by the thermal creation of
Sn vacancies \cite{DHBR16}. In the study by Dewandre \etal, the standard
semiconductor behavior discussed above, i.e.\ the thermal excitation of
electrons from the valence to the conduction band states and the
associated intrinsic, bipolar conductivity were not considered. Instead,
the formation energies of a number of intrinsic defects (vacancies,
interstitial atoms, Sn/Se exchange) were determined in DFT-based
calculations. The activation energy of the lowest-energy defect, the Sn
vacancy ($V^{(2-)}_\text{Sn})$ was found to be close to an activation
energy (0.67~eV) that describes the variation in carrier concentration
between 600 and 800~K, as deduced from the inverse Hall coefficient by
Zhao \etal\ \cite{ZLZS14} (Fig.~\ref{fig:invHall}). Based on the
similarity of the activation energy and Sn$^{2-}$ vacancy formation
energy, the rise in carrier concentration between 600 and 800 K was
attributed to the creation of additional Sn vacancies.

We would like to point out that in a bulk semiconductor, the creation of a Sn
vacancy ($\text{V}_\text{Sn}$) requires the creation of at least one secondary
defect in order to conserve the bulk stoichiometry. For example, the Sn ion
removed during the creation of a Sn vacancy could move to an interstitial site,
creating a secondary interstitial defect (Sn$_\text{Int}$), or occupy a Se site
together with Se moving to an interstitial site ($\text{Sn}_\text{Se} +
\text{Se}_\text{Int}$). Consequently, the effective activation energy will be
substantially higher than for the Sn vacancy alone, which is by far the
lowest-energy defect considered in \cite{DHBR16}, and the creation of free
charge carriers by this mechanism will be exponentially suppressed compared to
the results by Dewandre \etal\ (reproduced in Fig.~\ref{fig:invHall}). Even
though thermal activation of defects will undoubtedly occur at sufficiently
high temperatures with respect to the effective activation energies, we
conclude that this mechanism can neither explain the observed complex variation
of the electrical conductivity (or free carrier concentration) with
temperature, nor is it necessary.


Figure~\ref{fig:sigma-Seebeck}(b) shows the calculated Seebeck coefficient as a
function of temperature for temperature gradients along the three crystal
directions. The results for the directions $b$ and $c$ parallel to the crystal
layers ($S_{yy}, S_{zz}$) are in excellent agreement with the experimental
results by Zhao \etal\ \cite{ZLZS14} between 300 and 800~K. At higher
temperatures, the calculations yield a continued gradual decrease of the
Seebeck coefficient with increasing temperature, whereas experimental studies
found the Seebeck coefficient to be temperature-independent above 800~K
\cite{ZLZS14,ZWSQ17}.  A possible explanation for this and the corresponding
deviation for the conductivity is that the band gap assumes a minimum value at
$\sim$800~K and increases slightly thereafter up to 1000~K, similar to what we
observed by suppressing the small anomaly in the Sn-$\Gamma1$ distortion mode
[Fig.~\ref{fig:DFT_gap_dEtot}(a)]. Such an effect may well arise from the
anharmonic atomic motion, but it would require extensive modeling to test this
hypothesis. Unlike the electrical conductivity, the Seebeck coefficient and its
temperature dependence are sensitive to details of the band structure beyond
the band gap \cite{SupplMat} as discussed previously \cite{KWKT15,SK15}.

The computational results for $S_{xx}$, in the direction perpendicular to the
layers, deviate from the experimental results by Zhao \etal\ \cite{ZLZS14}. We
obtained a much more pronounced reduction of the Seebeck coefficient and a
change in sign at 680~K. Around 800~K, the calculations yield a large negative
Seebeck coefficient of $S_{xx} \approx \SI{-400}{{\mbox{\textmu}}V/K}$, i.e.\ with a larger
absolute value than for the $b$ and $c$ directions. Similar results were also
obtained in other recent computational studies \cite{KWKT15,SK15}, but their
relevance (see below) was perhaps not fully appreciated. As discussed
previously \cite{KWKT15,SK15}, the negative sign of $S_{xx}$ is a result of
bipolar conductivity in SnSe above 700~K. The magnitude of the band gap is
therefore an important factor \cite{WXTJ17}, and the agreement of our
calculations with the experimental results for the electrical conductivity and
inverse Hall coefficient and their temperature dependences indicates that the
temperature-dependent band gap used here is realistic. The result of a negative
$S_{xx}$ is a rather robust outcome of several calculations, despite
significant variations with respect to the band shapes and band gaps, and this
result should not simply be rejected because it is at variance with the
experimental results by Zhao \etal\ \cite{ZLZS14}. In fact, in a recent
single-crystal transport study up to \SI{700}{K} by Ibrahim \etal\
\cite{IVSC17}, a rapid reduction in $S_{xx}$ above 500~K was reported, and the
data are suggestive of a change in sign at around \SI{750}{K}
[Fig.~\ref{fig:sigma-Seebeck}(d)]. And in a study on polycrystalline samples, a
rapid reduction of the Seebeck coefficient was observed above 600~K, decreasing
down to $\sim$\SI{145}{{\mbox{\textmu}}V K^{-1}} at 960~K \cite{SCVO14}. The polycrystalline
average of positive and negative Seebeck coefficients would explain such a low
value.

These experimental results not only lend support to our computational results,
they also highlight the limitations and inconsistencies of the currently
available experimental data. Our computational work is based on the assumption
that the extrinsic carrier concentration remains constant in the 300--1000~K
range, which implies that the acceptors are shallow impurities, as has
generally been assumed. A possible explanation for the differences in the
transport properties in different studies is the existence of deep acceptors in
some samples, which could lead to substantial differences in the extrinsic
carrier concentration and hence transport properties at high temperatures. A
re-examination of the single-crystal transport properties of SnSe above 700~K,
and also the influence of different acceptor types, would be extremely
valuable. Controlling the defect concentration and keeping it below
$\sim$\SI{E18}{cm^{-3}} would be crucial for observing the negative $S_{xx}$
Seebeck effect.

The prediction of a large Seebeck effect with opposite signs along two
directions of a bulk crystal may seem very unusual, but there are precedents
for such behavior: Early work on bismuth suggests a change in sign along one
crystal direction at 280{\textdegree}C \cite{Boy27}, and nominally undoped, $p$-type
\ce{Bi2Te3} has been reported to show Seebeck coefficients with opposite signs
between 240 and 390{\textdegree}C, which was attributed to bipolar conductivity
\cite{Den61}.

The simultaneous occurrence of a positive and negative Seebeck effect in highly
anisotropic, layered materials at the onset of bipolar conduction is an
interesting new aspect that could be exploited in thermoelectric applications.
It would eliminate the need to find suitable dopants to produce both $p$- and
$n$-type versions of a given compound. Instead, a single, highly anisotropic
material could be used in different orientations, i.e. parallel and
perpendicular to the layers. Such a material does not necessarily need to be
single-crystalline, a highly oriented polycrystalline material \cite{GXQZ16}
would be sufficient and preferable in terms of production cost and mechanical
properties.

\bigskip

\section{Conclusions}

Our re-examination of the structural phase transition in SnSe using
high-resolution powder neutron diffraction and symmetry-mode analysis
identifies Sn displacements parallel to the SnSe layers (\mbox{Sn-Y2}
mode) as a primary order parameter and provides evidence for anomalous Sn
displacements perpendicular to the rocksalt layers (Sn-$\Gamma$1 mode)
that have profound effect on calculated electronic properties. The
magnitude of the band gap is directly related to the \mbox{Sn-Y2} mode
amplitude during the transition. The magnitude of the calculated band gap
was found to be sensitive to the atomic coordinates so that simulations
based on DFT-relaxed crystal structures do not work sufficiently well in
SnSe --- a limitation that probably applies more generally to materials
with anharmonic, soft chemical bonds.

The transport property calculations provide a coherent explanation of the unusual
step-like variation of the electrical conductivity observed experimentally in SnSe
\cite{ZLZS14}. This behavior originates from the onset of intrinsic bipolar conductivity,
amplified and shifted to lower temperatures by the rapid reduction of the band gap
between 700 and 800~K. The calculated thermopower becomes highly anisotropic at
temperatures above 600~K, and SnSe is predicted show simultaneously large positive and
negative Seebeck coefficients along different crystal directions. This prediction is
consistent with one experimental study limited to 700~K \cite{SCVO14}, but at variance
with another \cite{ZLZS14}. On the experimental side, it may prove important to control
the concentration of deep defects in SnSe \cite{PC90,ACPL94} so as to avoid an increase
of the extrinsic carrier concentration above room temperature. It is worthwhile to
identify in future research highly anisotropic materials with simultaneously large
positive and negative Seebeck effects as a possible route towards more efficient
thermoelectric materials.

\acknowledgments SRP and JWGB acknowledge the EPSRC (EP/N01717X/1), Leverhulme
Trust (RPG-2012-576) and the STFC for provision of beam time \#1520262 at ISIS
\cite{ISIS-Exp:1520262}. Supporting data are available from the University of
Edinburgh{'}s Datashare repository at \url{http://datashare.is.ed.ac.uk},
DOI:\href{http://dx.doi.org/10.7488/ds/2401}{10.7488/ds/2401}.



\end{document}